\documentclass[12pt,a4paper]{article}

\usepackage{graphicx}
\usepackage{amssymb}

\newtheorem{lemma}{Lemma}
\newtheorem{prop}[lemma]{Proposition}

\newcommand{\be}{\begin{equation}}
\newcommand{\ee}{\end{equation}}

\newcommand{\bea}{\begin{eqnarray}}
\newcommand{\eea}{\end{eqnarray}}

\newcommand{\nn}{\nonumber}

\newcommand{\rh}{r_{h}}

\begin{document}
\begin{titlepage}
 
\begin{flushright}
OUTP--98--92--P \\
\end{flushright}
 
\begin{centering}
\vspace{.05in}
{\Large {\bf {Existence of stable hairy black holes in
    ${\mathfrak {su}}(2)$  Einstein-Yang-Mills theory 
with a negative cosmological constant
}}}
\vspace{.2in}

E. Winstanley \\
\vspace{.05in}
University of Oxford, Department of Physics,
 Theoretical Physics, 
1 Keble Road, Oxford. OX1 3NP, U.K. \\

\vspace{.2in}
{\bf Abstract} \\
\vspace{.05in}
\end{centering}
We consider black holes in EYM theory with a negative
cosmological constant.
The solutions obtained are somewhat different from those
for which the cosmological constant is either
positive or zero.
Firstly, regular black hole solutions exist for 
continuous intervals of the parameter space,
rather than discrete points.
Secondly, there are non-trivial solutions in which
the gauge field has no nodes.
We show that these solutions are linearly stable.
\vspace{2.5in}

\begin{flushleft} 
December 1998 \\
\end{flushleft} 

\end{titlepage}

\section{Introduction}
Non-Abelian Einstein-Yang-Mills (EYM) theories have been the subject of 
detailed study since the discovery of particle-like \cite{bartnik}
and black hole \cite{bizon} solutions when the gauge group is 
${\mathfrak {su}}(2)$.
Since then, there has been a great deal of numerical and analytic
work on various aspects 
of ${\mathfrak {su}}(N)$ EYM black holes and solitons, in both
asymptotically flat and asymptotically de Sitter geometries
(see \cite{review} for a wide-ranging, recent review 
of work to date, and \cite{sun} for some analytic 
results). The behaviour of the cosmological solutions is similar in
many respects to that for asymptotically flat geometries, for
sufficiently small, positive, 
cosmological constant \cite{volkov,coscol} and in
particular, the configurations are unstable \cite{cosstab}.
Our main result is that there are black hole
solutions of ${\mathfrak {su}}(2)$ EYM theory with a
negative cosmological constant which 
 are {\em {stable}}.

\bigskip
In this paper we are concerned with  EYM 
black holes in spacetimes which are asymptotically anti-de Sitter (AdS).
AdS space has recently enjoyed a revival due to the work of
Witten and others in connection with conformal field theories and
large $N$ gauge theories (see, for example, \cite{witten}).
Initial interest in black holes that are asymptotically AdS was 
due to the result that (sufficiently large)
Schwarzschild-anti-de Sitter black holes are
thermodynamically stable \cite{hawking}, which has been the
inspiration for much of the recent work.
Another trigger was the discovery of three-dimensional black holes
constructed from AdS space \cite{btz}, which have interesting thermodynamical
properties.  
This suggests that, although geometries with a negative cosmological 
constant have some undesirable properties (such as closed timelike curves,
although these are readily removed by considering the covering space)
it has many advantages for studying quantum field theory and may prove
to be a useful probe of quantum gravity effects.
Black holes with hair in such theories may therefore be useful for
probing not only quantum gravity, but also may be interesting tests
of the AdS-CFT conjecture \cite{maldacena}, particularly since
we can find such objects which are classically stable.

\bigskip
Here we restrict attention to ${\mathfrak {su}}(2)$ EYM theory for 
simplicity, and investigate black hole solutions which
asymptotically approach AdS space.
Numerical calculations will be backed up with analytic work.
We find some surprising results, which are rather different from the
corresponding ones for asymptotically flat and de Sitter black holes,
of which stability of some of our solutions is the most important. 

\bigskip
The outline of the paper is as follows. 
Firstly, in section \ref{field}
we outline the field equations and ansatze for the 
field variables we shall be using, before discussing the numerical 
solution of these field equations, integrating out from
the event horizon.
The results found are rather 
different from those observed in the case of a positive cosmological 
constant \cite{volkov,coscol}. 
For example, here the solutions are distributed continuously in the 
space of initial parameters, rather than discretely, and there are 
solutions in which the gauge field has no zeros.
These numerical results are backed up with analysis of the
defining differential equations in section \ref{analytic}.
This section emphasizes the differences in the behaviour of the
field equations when the cosmological constant is positive,
negative, or zero. 
Those solutions for which the gauge field has no zeros are
particularly interesting since they are
shown to be stable, by a mixture of analytic and numerical 
techniques, in section \ref{stable}.
Finally, our conclusions are presented in section \ref{conc}.

\bigskip
The metric signature is $(-+++)$ throughout, and we use units in
which $G=1=c$.

\section{Solving the field equations}
\label{field}

\subsection{Preliminaries}

The Einstein equations with a cosmological constant 
$\Lambda $ take the form
\be
R_{\mu \nu }-\frac {1}{2} R g_{\mu \nu } +\Lambda g_{\mu \nu }
=8 \pi T_{\mu \nu }.
\ee
For simplicity we restrict attention to spherically  symmetric
solutions of these equations, in the case that $\Lambda <0$.
We require the geometry to approach (the covering space of)
AdS spacetime at infinity, and
therefore a suitable ansatz for the metric is
\bea
ds^{2}& = &  -\left( 1-\frac {2m(r)}{r}-\frac {\Lambda r^{2}}{3} \right)
S^{2}(r) dt^{2} +
\left( 1-\frac {2m(r)}{r}-\frac {\Lambda r^{2}}{3} \right) ^{-1}
dr^{2} \nn \\
& & 
+r^{2} \left( d\theta ^{2}+\sin ^{2}\theta \, d\phi ^{2}
\right) ,
\eea
where $m(r)$ will be the `quasi-local mass' for the
geometry relative to anti-de Sitter spacetime itself.
The most general static, spherically symmetric ansatz for the 
${\mathfrak {su}}(2)$ Yang-Mills gauge field, with a purely
magnetic field and 
an appropriate choice of gauge, is 
\be
A=(1+\omega )\left[ -\tau _{\phi }\, d\theta +\tau _{\theta }
\sin \theta \, d\phi \right]
\label{staticansatz}
\ee
where $\omega $ is a function of $r$ alone, and the 
$\tau _{r,\theta ,\phi }$ are given in terms of the usual
Pauli matrices $\tau _{1,2,3}$ by
$\tau _{r, \theta ,\phi }=
{\mbox {\boldmath {$\tau .e_{r,\theta ,\phi }$}}} $.

\bigskip
The theory has, in addition to Newton's constant
$G$ (which we have set equal to unity by an appropriate
choice of units), a gauge coupling constant $g$
and cosmological constant $\Lambda $, all of which 
are parameters.
We shall set the gauge coupling constant $g$ also equal
to unity, leaving $\Lambda $ as the only parameter.
This is the approach taken by the authors of ref. 
\cite{volkov}.
In \cite{coscol} a different combination of coupling
constants was used as the single free parameter.

\bigskip
The field equations then take the
following form, with $'$ denoting $d/dr$:
\bea
m' & = & 
\left( 1-\frac {2m(r)}{r}-\frac {\Lambda r^{2}}{3} \right)
\omega '^{2} +
\frac {\left( \omega ^{2}-1 \right) ^{2}}{2r^{2}} 
\label{first}
\\
\frac {S'}{S} & = & 
\frac {2\omega '^{2}}{r} 
\label{Seqn}
\\
0 & = & 
r^{2} \left( 1-\frac {2m(r)}{r}-\frac {\Lambda r^{2}}{3} \right)
\omega '' 
+\left( 2m-\frac {2\Lambda r^{3}}{3}-\frac {\left( \omega ^{2}-1
\right) ^{2}}{r} \right) \omega '
\nn \\ & & 
+\left( 1-\omega ^{2}\right) \omega .
\label{last}
\eea
These equations are invariant under the transformation 
$\omega \rightarrow -\omega $, although the ansatz 
(\ref{staticansatz}) is not.
The gauge potentials (\ref{staticansatz}) for 
$\pm \omega $ are related by a gauge transformation of the 
form \cite{review}
\be
A\rightarrow UAU^{-1}+UdU^{-1},
\qquad
U=\exp \left[ \pi \tau _{r} \right] .
\ee
Since we are interested in black hole solutions, there
are two length scales to consider,
firstly, the event horizon radius $r_{h}$, and secondly
the length scale $l^{-2}=-\frac {\Lambda }{3}$.
For simplicity, in our numerical work in section 
\ref{numerical}, we shall set $r_{h}=1$ and vary
$\Lambda $ in order to change the relative magnitude
of the two length scales.
However, the analytic results that follow will hold for all
$r_{h}$.

\bigskip
For black hole solutions having a regular event horizon at
$r=\rh $, 
\be
m(\rh )=\frac {\rh }{2}-\frac {\Lambda \rh ^{3}}{6}>0
\qquad \forall \rh >0
\label{mh}
\ee
since $\Lambda <0$.
In order to have a regular event horizon at $r=\rh $, it must be
the case that
\be
\left.
\frac {d}{dr} \left( 
1-\frac {2m}{r} -\frac {\Lambda r^{2}}{3} \right) 
\right| _{r_{h}} >0.
\ee
This places a bound on $m'(\rh) $:
\be
2m'(\rh ) =\frac {\left[ \omega (\rh )^{2}-1 \right] ^{2}}{\rh ^{2}}
<1-\Lambda \rh ^{2} .
\label{mhbound}
\ee
From the field equations, for fixed $r_{h}$,
there is just one initial parameter, 
$\omega _{h}=\omega (\rh)$, 
since then $\omega '(\rh )$ is given by
\be
\omega '(\rh )=\frac {\left[ \omega _{h}^{2}-1 \right] 
\omega _{h}}{\rh -\Lambda \rh ^{3}-\frac {\left[
\omega _{h}^{2}-1\right] ^{2}}{\rh }} .
\label{omegah}
\ee 
Since the cosmological constant is negative, we do not expect there
to be a cosmological event horizon.  
This places  a bound on $m(r)$, namely,
\be
m(r)< \frac {r}{2}-\frac {\Lambda r^{3}}{6}.
\label{mbound}
\ee
This is quite a weak restriction, whereas when $\Lambda $ is positive,
there will always be a cosmological horizon, and fairly strict
conditions have to be placed on the field variables at the
cosmological horizon in order that it is regular.
When $\Lambda =0$, the restriction (\ref{mbound})
is stronger than for $\Lambda <0$, and will be violated 
for generic initial data \cite{bfm}.
As $r\rightarrow \infty $,  the solutions are expected to be charged
as in the $\Lambda >0$ case \cite{volkov}, and the field  variables will
have the following asymptotic forms:
\bea
m(r) & = & M+\frac {M_{1}}{r} +O\left( \frac {1}{r^{2}} \right) \nn \\
\omega (r) & = & \omega _{\infty } +\frac {a}{r} +
\frac {b}{r^{2}}+O\left( \frac {1}{r^{3}} \right) \nn \\
S(r) & = & 1+O\left( \frac {1}{r} \right) ,
\label{inf}
\eea
where  $\omega _{\infty }$, $M$ and $a$ are parameters and
\bea
M_{1} & = & \frac {\Lambda a^{2}}{3}-\frac {1}{2}\left( 
\omega _{\infty }^{2}-1 \right) ^{2} \nn \\
b & = & \frac {3\omega _{\infty }\left( 1-\omega _{\infty }\right)
}{2\Lambda } .
\label{infconds}
\eea

\subsection{Numerical Results}
\label{numerical}
The field equations (\ref{first}--\ref{last}) were integrated
numerically, fixing $\rh =1$, so that $\Lambda $ was the only
parameter in the problem, and $l$ 
(where $l^{-2}=-\frac {\Lambda }{3}$) is the only 
varying length scale.  
Using the initial conditions on the event horizon
(\ref{mh},\ref{omegah}), the equations were integrated for increasing
$r$, for various values of $|\Lambda |$ in the range
$10^{-4}$--$10^{6}$, and a range of values of $\omega _{h}$.
Since the equations (\ref{first}-\ref{last}) are invariant under the
transformation $\omega \rightarrow - \omega $, only values of
$\omega _{h}>0$ were considered.  
The equation for $S$ decouples from the rest, and 
the requirement that $S\rightarrow 1 $ as $r\rightarrow \infty $
can be relaxed during the numerical integration, 
$S$ subsequently being multiplied by an appropriate constant factor so
that the correct asymptotic behaviour is recovered.
Some of the results found were rather surprising, and are listed below.

\begin{figure}
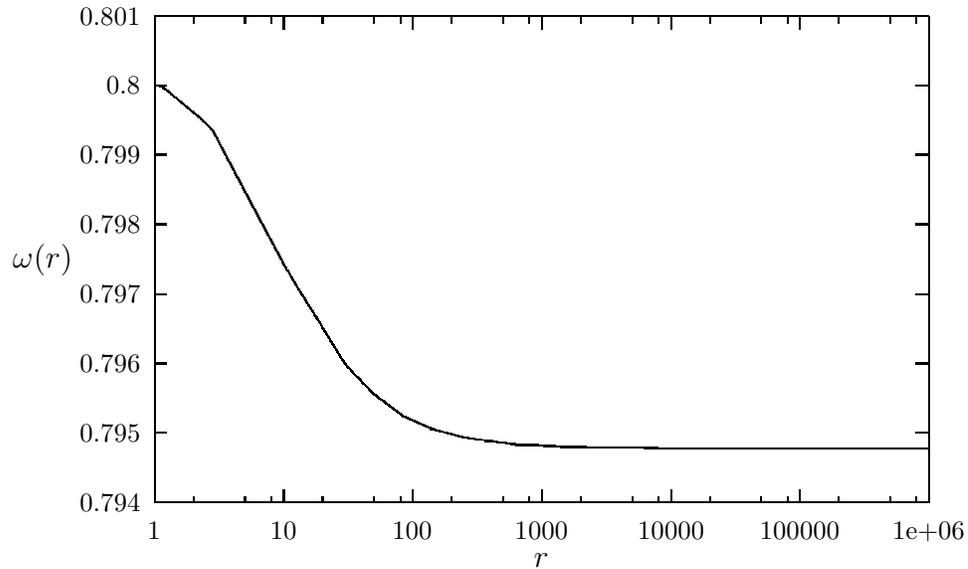

% GNUPLOT: LaTeX picture
\setlength{\unitlength}{0.240900pt}
\ifx\plotpoint\undefined\newsavebox{\plotpoint}\fi
\sbox{\plotpoint}{\rule[-0.200pt]{0.400pt}{0.400pt}}%

\caption{Black hole solution for which $\omega $ has no nodes, with 
$\Lambda =-100$ and $\omega _{h}=0.8$.}
\label{l100w08}
\end{figure}

\begin{figure}
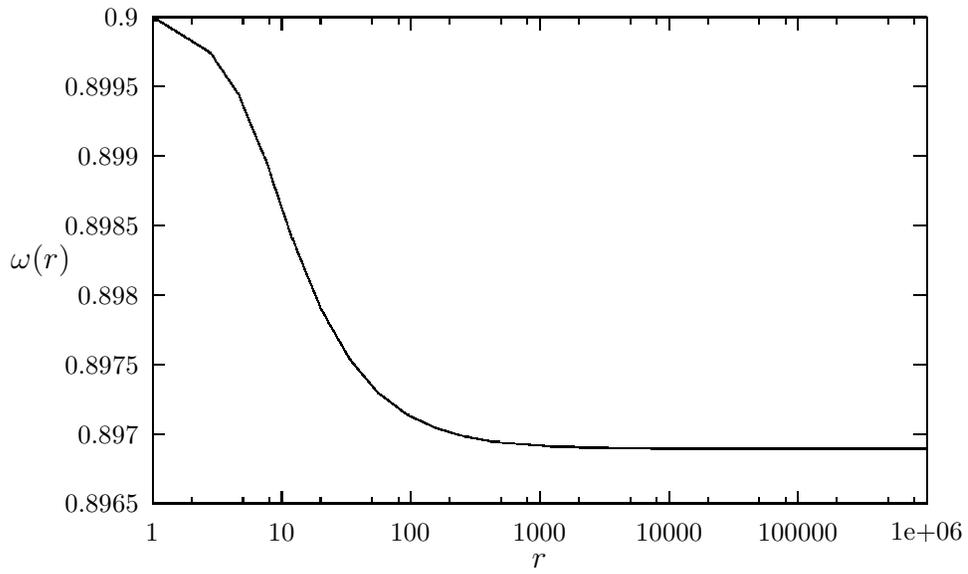

% GNUPLOT: LaTeX picture
\setlength{\unitlength}{0.240900pt}
\ifx\plotpoint\undefined\newsavebox{\plotpoint}\fi
%
\caption{As figure \ref{l100w08}, but with $\omega _{h}=0.9$.}
\label{l100w09}
\end{figure}

\begin{figure}
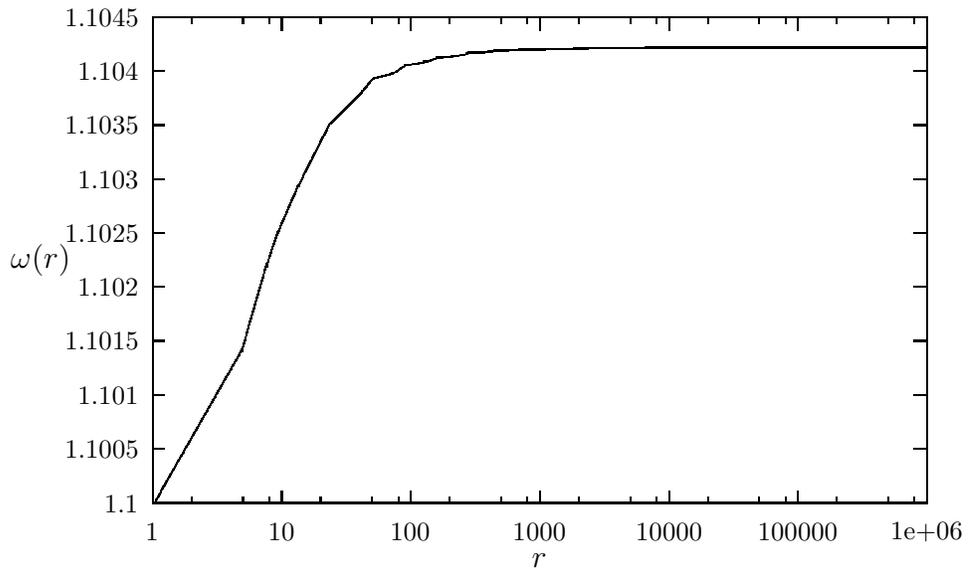

% GNUPLOT: LaTeX picture
\setlength{\unitlength}{0.240900pt}
\ifx\plotpoint\undefined\newsavebox{\plotpoint}\fi
%
\caption{Black hole solution in which $|\omega |>1$ for all $r$.
Here $\Lambda =-100$, and $\omega _{h}=1.1$.}
\label{l100w11}
\end{figure}

\begin{figure}
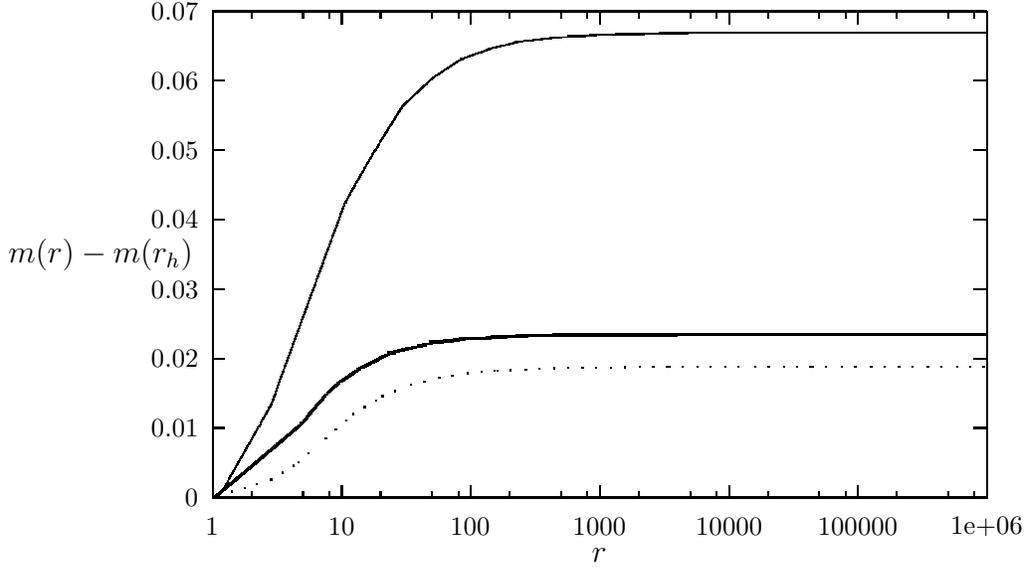

% GNUPLOT: LaTeX picture
\setlength{\unitlength}{0.240900pt}
\ifx\plotpoint\undefined\newsavebox{\plotpoint}\fi
\sbox{\plotpoint}{\rule[-0.200pt]{0.400pt}{0.400pt}}%
%
\caption{Black hole solutions in which the gauge field $\omega $
has no nodes, for $\Lambda =-100$.
Here we plot $m(r)-m(\rh )$, with $m(\rh )=103/6$ for this value
of $\Lambda $.
In this figure, and also figures \ref{l100d} and \ref{l100u}, 
the solid line represents the solution for $\omega _{h}=0.8$, the
dotted line $\omega _{h}=0.9$ and the bold solid line 
$\omega _{h}=1.1$.}
\label{l100m}
\end{figure}

\begin{figure}
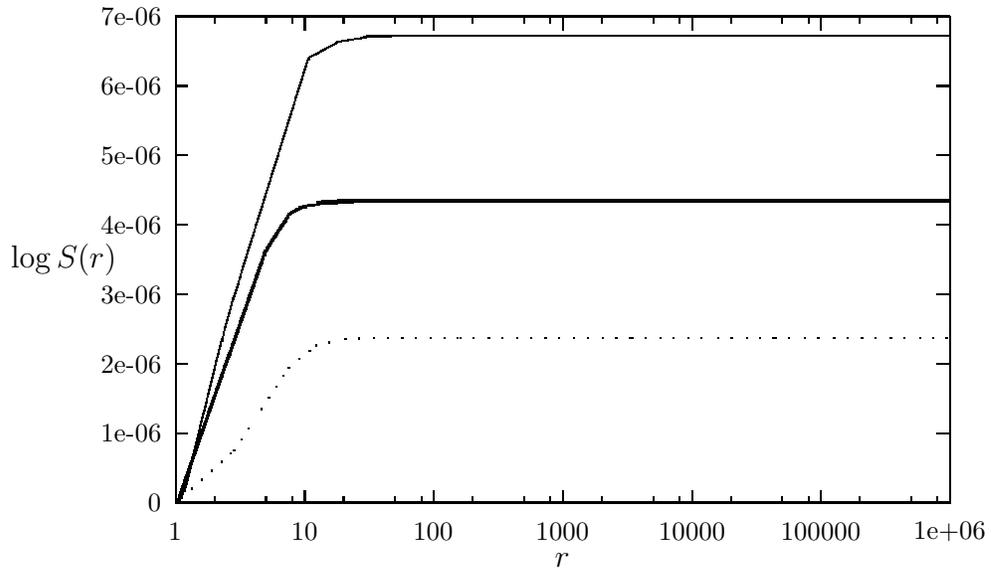

% GNUPLOT: LaTeX picture
\setlength{\unitlength}{0.240900pt}
\ifx\plotpoint\undefined\newsavebox{\plotpoint}\fi
%
\caption{The same solutions as figure \ref{l100m}, but here
$\log S (r)$ is plotted.}
\label{l100d}
\end{figure}

\begin{figure}
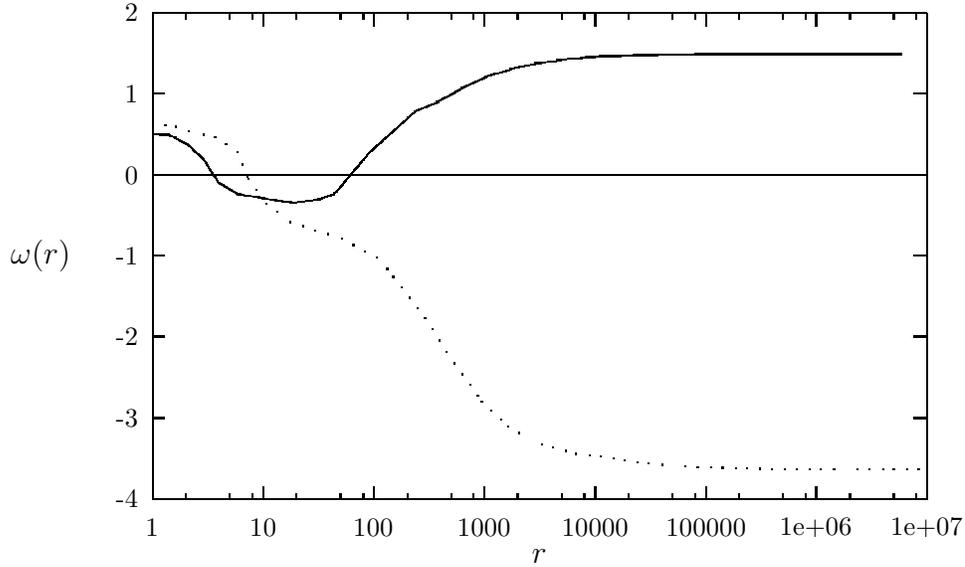

% GNUPLOT: LaTeX picture
\setlength{\unitlength}{0.240900pt}
\ifx\plotpoint\undefined\newsavebox{\plotpoint}\fi
\sbox{\plotpoint}{\rule[-0.200pt]{0.400pt}{0.400pt}}%
%
\caption{Black hole solutions in which $|\omega (r)|>1$ for $r$
sufficiently large, although $|\omega (\rh )|<1$.
Here, and in figures \ref{l001m} and \ref{l001d},
we use cosmological constant
$\Lambda =-0.001$.
The solid line represents the solution
with $\omega _{h}=0.5$ (the gauge field having 2 nodes) and
the dotted line corresponds to $\omega _{h}=0.64$ (1 node).
}
\label{l001w}
\end{figure}

\begin{figure}
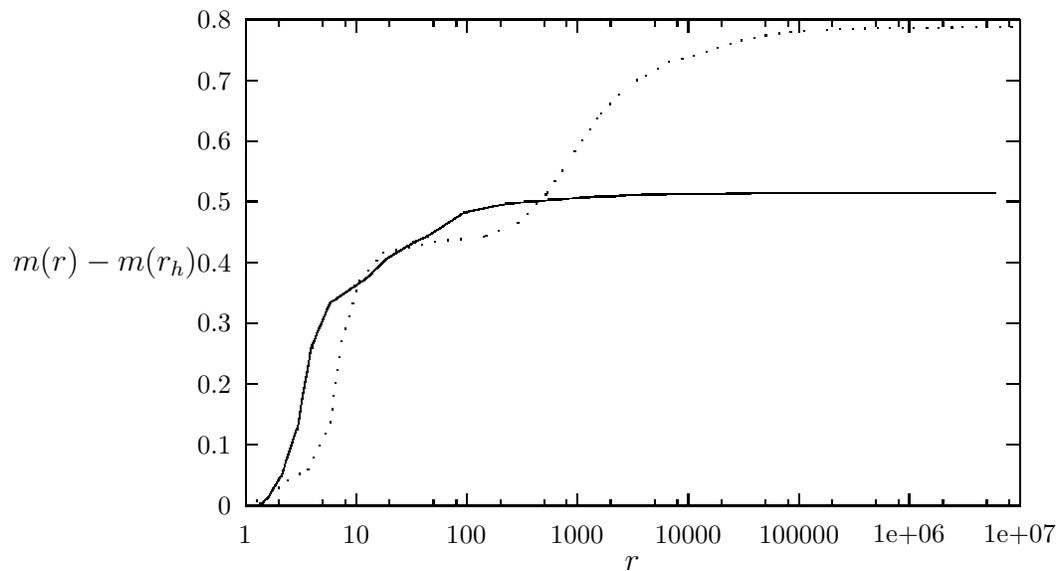

% GNUPLOT: LaTeX picture
\setlength{\unitlength}{0.240900pt}
\ifx\plotpoint\undefined\newsavebox{\plotpoint}\fi
%
\caption{The metric function $m(r)$ plotted for the solutions of 
figure \ref{l001w}. For this value of $\Lambda $, note that 
$m(\rh )\simeq 0.5$.}
\label{l001m}
\end{figure}

\begin{enumerate}
\item
For $|\Lambda |$ sufficiently large (i.e. $\ge 0.1$)
there exist regular solutions for which the gauge field has no zeros
(nodes).
Examples of this type of solution are illustrated in figures
\ref{l100w08} to \ref{l100d}, for the value $\Lambda =-100$.
For this large value of $|\Lambda |$, the field variables do not
vary much as $r$ increases. 
This is in accordance with proposition \ref{largeprop} in the
next section.
This is the first example of behaviour unique to the
negative cosmological constant theory. 
For $\Lambda \ge 0$, the gauge field must have at least one
zero.
\item
For every value of $\Lambda <0$, there was a critical value,
$\omega _{h}^{c}>0$, of $\omega _{h}$,
such that, for all $0<\omega _{h}<\omega _{h}^{c}$ 
integrating the field equations gave a regular black hole solution
which was asymptotically AdS.
The value of $\omega _{h}^{c}$ increases as 
$|\Lambda |$ increases.
For $|\Lambda |$ sufficiently large (i.e. $\ge 0.1$), 
$\omega _{h}^{c}>1$.
Figure \ref{l100w11} illustrates the solution when
$\Lambda =-100$ and $\omega _{h}=1.1$. 
When $\Lambda \ge 0$, the regular solutions occur only for
discrete values of the initial parameter, although these
values do have an accumulation point at zero.
For $\Lambda =0$, it has been shown \cite{bfm} that
it is not possible to have regular black holes for which
$|\omega (r)|>1$ for any $r$, and all evidence to date shows
that we may expect this to hold also for positive cosmological
constant \cite{volkov,coscol}.
\item
Unlike the situation when $\Lambda \ge 0$,
the regular black hole solutions did not occur for discrete values
of $\omega _{h}$, but, even for $\omega _{h}>\omega _{h}^{c}$,
over continuous intervals of $\omega _{h}$.
For smaller values of $|\Lambda |$ ($\sim 10^{-4}$), these 
intervals were separated by intervals on which there were no
regular solutions.
For each $\Lambda <0$, there is a second, maximal, value
of $\omega _{h}$, $\omega _{h}^{m}>\omega _{h}^{c}$,
such that for all $\omega _{h}>\omega _{h}^{m}$
there are no regular black hole solutions.
This behaviour is also noticed for $\Lambda \ge 0$,
and follows from the condition (\ref{mhbound}) for
the existence of a regular event horizon.
\item
There are regular solutions for which $|\omega _{\infty}|>1$,
although $|\omega _{h}|<1$.
Examples of this type of solution can be seen in figures
\ref{l001w} to \ref{l001d}, when $\Lambda =-0.001$.
Again, when $\Lambda \ge 0$, this type of solution is 
forbidden since $|\omega (r)|<1$ for all $r$ inside the 
cosmological event horizon.
\item
As $|\Lambda |$ increases, the values of $\omega _{h}$ for which the
regular solutions have one or more nodes decrease rapidly. 
\end{enumerate}
Otherwise, many of the properties of the numerical solutions
were as found in the asymptotically flat \cite{bizon} or 
asymptotically de Sitter \cite{volkov,coscol} models:
\begin{enumerate}
\item
For each regular black hole solution extending to 
$r\rightarrow \infty $, the gauge field $\omega $ has a finite number
of zeros.
\item
The number of zeros of $\omega $ increases as $\omega _{h}$ decreases.
\end{enumerate}

\section{Analytic explanation of the numerical results}
\label{analytic}

In this section we prove some analytic results which explain the 
behaviour observed numerically in the previous section.
We shall stress, where appropriate, the similarities and differences
between our results and the corresponding ones for asymptotically
flat and asymptotically de Sitter spacetimes.
In the results of this section, we shall pay little attention to the
behaviour of the metric function $S(r)$.
The regularity of this function can easily be established in each case
by the appropriate behaviour of the other field variables, and
integrating its defining equation (\ref{Seqn}), 
with the inclusion of a constant (see discussion in the previous
section).

\subsection{Elementary results}
We begin with some simple lemmas, the proofs of which are identical
to those in the $\Lambda =0$ case \cite{bfm}, and which are valid
for all values of $\Lambda $.

\begin{lemma}
\label{horexist}
For each fixed value of $\Lambda $, there exists a 
family of local solutions of the field equations
(\ref{first}-\ref{last}) satisfying the initial conditions 
(\ref{mh},\ref{omegah}), defined for $\rh >0$, and $\omega _{h}$
such that (\ref{mhbound}) is satisfied, and analytic in $\rh$, 
$\omega _{h}$, $r$ and $\Lambda $.
\end{lemma}

\begin{lemma}
For each value of $\Lambda $, there exists a 
family of local solutions of the field equations 
(\ref{first}-\ref{last}) satisfying the boundary conditions
(\ref{inf},\ref{infconds}), and analytic in $1/r$, 
$\omega _{\infty }$, $a$, $M$, and $\Lambda $.
\end{lemma}

\begin{lemma}
\label{reglem}
As long as 
\be
1-\frac {2m(r)}{r} -\frac {\Lambda r^{2}}{3}>0,
\ee
the solutions are regular functions of $r$.
\end{lemma}

\noindent
Note that lemma \ref{reglem} only guarantees regularity of the
solutions having a positive cosmological constant up to the 
cosmological event horizon.  
Conditions (similar to those for a black hole event horizon) have
to be imposed on the field variables if the cosmological horizon
is to be regular.  
It is the necessity of these conditions which makes the behaviour
of the solutions of the Einstein-Yang-Mills equations with
positive cosmological constant so different from those we discuss
here with negative cosmological constant.

\begin{figure}
\begin{center}
\includegraphics[angle=270,width=10cm]{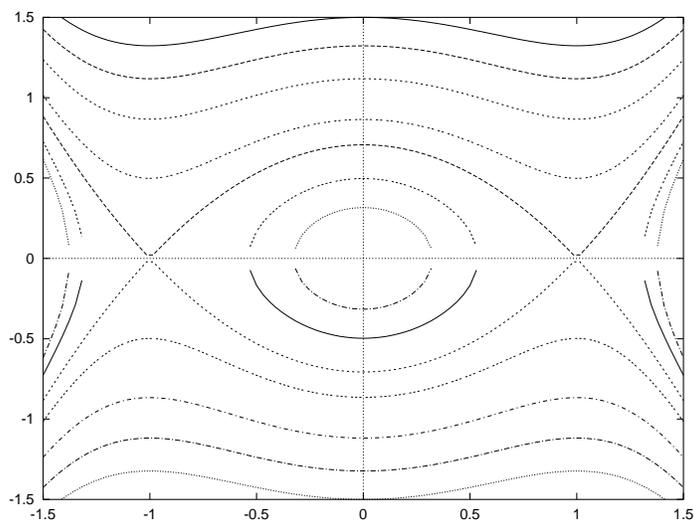}
\end{center}
\caption{The phase plane for the equation 
$
\frac {d^{2}\omega }{d\tau ^{2}} =\left( \omega ^{2}-1\right) \omega .
$
}
\label{phase}
\end{figure}

\bigskip
The argument in the next subsection will make use of the asymptotic 
solutions of the field equations as $r\rightarrow \infty $.
Therefore we consider the Yang-Mills equation in the regime
$r\gg 1$, $|\Lambda |r^{2}\gg 1$, which takes the form:
\be
-\frac {\Lambda r^{4}}{3}\omega ''-\frac {2\Lambda r^{2}}{3}
\omega ' +\left( 1-\omega ^{2} \right) \omega =0.
\label{asympteqn}
\ee
This equation is made autonomous by the use of the change of variable
\be
\tau = \frac {l}{r},
\label{taudef}
\ee
where $l^{-2}=-\frac {\Lambda }{3}$.
Then the equation is,
\be
\frac {d^{2}\omega }{d\tau ^{2}} =\left( \omega ^{2}-1\right) \omega .
\label{asympt}
\ee
The phase portrait of this equation is shown in figure \ref{phase}.
For $\Lambda >0$, the right-hand side of equation
(\ref{asympt}) has the opposite sign, which means that the phase
plane is rather different. 
However, this is less crucial than the inevitability of
a cosmological event horizon in this case.
The phase plane for the asymptotically flat case in which
$\Lambda =0$ is discussed in \cite{bfm}.
In order to make the equation corresponding to 
(\ref{asympteqn}) in that case autonomous, the variable
has to be changed to 
${\tilde {\tau }}=\log r$, rather than (\ref{taudef}).
This means that as $r\rightarrow \infty $, the new variable
${\tilde {\tau }}\rightarrow \infty $ also, whereas in our
case $\tau \in [0,\tau _{1}]$ for very large $r$, as 
$r\rightarrow \infty $, in other words, $\tau $ remains in
a finite interval.  
This is responsible for the surprising results of propositions
\ref{smallprop}-\ref{largeprop} below, since phase
paths that are close to the saddle points in figure
\ref{phase} do not necessarily have to travel off to 
infinity as $r\rightarrow \infty $, since $\tau $ remains
finite.
However, for asymptotically flat black holes, the
phase portrait also has saddle points at 
$\omega =\pm 1$, and $\omega '=0$ (together with a 
spiral point at $\omega =0=\omega '$ rather than a centre),
but paths passing close to the saddle points must zoom
off to infinity because ${\tilde {\tau }}$ cannot remain
bounded as $r$ increases.

\subsection{Existence of a continuum of asymptotically AdS black
  holes}
\label{continuum}
The propositions in this section
 make use of the asymptotic equation
(\ref{asympt}) and the reader may ask whether they still apply
when $\Lambda >0$.
The answer is no, because we can no longer guarantee that if a 
solution has a regular cosmological event horizon, then integrating
the field equations with sufficiently close initial parameters will
also yield a solution with a regular cosmological horizon.
Indeed, one may suspect that in general this will not be the case.
In asymptotically flat spacetime \cite{bfm}, all solutions
close to the regular black holes are in fact singular 
because phase paths close to the saddle points must reach 
infinity, as discussed in the previous subsection.

\begin{prop}
For fixed $\rh $ and $\Lambda <0$, and for   
every $\omega _{h}$ sufficiently small, there is a regular,
asymptotically AdS black hole solution.
\label{smallprop}
\end{prop}

\noindent
{\bf{Proof}}
\smallskip
\newline
If $\omega _{h}=0$, then $\omega '(\rh )=0$ and we have the
Reissner-Nordstrom AdS (RNAdS) solution.
From the local existence theorem (Lemma \ref{horexist}), the 
solutions depend continuously on $\omega _{h}$ and therefore,
for sufficiently small $\omega _{h}$,
all field variables remain close to their values for the
RNAdS solution with the same $\rh $ until $r\gg 1$
and $ 2m(r)/r\ll 1$. 
In this situation the asymptotic form of the Yang-Mills equations
(\ref{asympt}) is valid.
The phase plane of this equation possesses a centre at
$\omega =0=\frac {d\omega }{d\tau }$, and therefore $\omega $ and
$\omega '=-\frac {1}{r^{2}l}\frac {d\omega }{d\tau }$ 
will remain very small
for all values of increasing $r$.
Therefore $m(r)/r$ will also continue to be extremely small for all $r$
and we have a regular black hole solution.
\hfill
$\square $

\bigskip
When $\Lambda =0$, the proof breaks down because 
$\omega $ and $\omega '$ no longer remain small as $r$
increases \cite{bfm}, and the solution will in general
become singular.

\begin{prop}
\label{contprop}
For fixed $\rh $ and $\Lambda <0$, if ${\bar {\omega }}_{h}$ leads to a 
regular black hole solution with the gauge field having $n$ nodes,
then all $\omega _{h}$ sufficiently close to ${\bar {\omega }}_{h}$
will also lead to a regular black hole solution with $n$ nodes.
\end{prop}

\noindent
{\bf{Proof}}
\smallskip
\newline
Since the solutions are continuous in $\omega _{h}$
from Lemma \ref{horexist}, we can choose $r_{1}\gg 1$ such that
$|\Lambda |r_{1}^{2}\gg 1$ and
for all $\omega _{h}$ sufficiently close to ${\bar {\omega }}_{h}$,
the gauge field function $\omega $ has $n$ nodes in the interval
$\rh < r < r_{1}$ and $m(r_{1})/r_{1}\ll 1$.
Therefore, as in the proof of the previous proposition, the 
asymptotic equation (\ref{asympt}) is applicable, where
$\tau =\tau _{1}\ll 1$.
If we have chosen $r_{1}$ sufficiently large, the field
variables $\omega $ and $\frac {d\omega }{d\tau }$ 
will not move very far on 
the appropriate phase plane trajectory (figure \ref{phase}) as
$\tau $ decreases from $\tau _{1}$ to zero.  
In particular, $\omega $ will not cross zero again, provided $r_{1}$
has been chosen to be sufficiently large.
In this case $m(r)/r$ will continue to be very small and this
will not affect the Yang-Mills equation.
Therefore we have a regular black hole solution with the gauge field
having $n$ nodes.
\hfill
$\square $

\bigskip
Here again, when $\Lambda =0$,
since $\omega $ and $\omega '$ will grow
indefinitely as $r$ increases, we can no longer guarantee
that a regular solution will be formed for $\omega _{h}$
sufficiently close to ${\bar {\omega }}_{h}$.
In general, a singular solution will be formed \cite{bfm}.

\subsection{Solutions for $|\Lambda |\gg 1$}

We close this section by proving the existence of regular black hole
solutions for which the gauge field has no nodes.  
These solutions will be particularly important as we shall show in the
next section that they are (linearly) {\em {stable}}.
Note that for $\Lambda =0$, it has been proved \cite{bfm}
that the gauge field must have a least one node.
This also appears to be the case for $\Lambda >0$
\cite{volkov,coscol}, although
there is more limited analytic work in this area \cite{linden}.

\begin{prop}
For fixed $\rh $,  given any value of $\omega _{h}$, 
for $|\Lambda |$ sufficiently large,
there exists a regular black hole solution in which $\omega $ has
no nodes.
\label{largeprop}
\end{prop}

\noindent
{\bf{Proof}}
\smallskip
\newline
First note that for all $|\Lambda |$ sufficiently large, the condition
(\ref{mhbound}) for a regular black hole event horizon at $r=\rh $
is satisfied.
The condition (\ref{mh}) for a regular event horizon suggests that
$m\sim |\Lambda |$ as $|\Lambda |\rightarrow \infty $.
In the light of this, define a new variable $p$ by
$p(r)=m(r)/\Lambda $ which will then remain finite as
$|\Lambda |\rightarrow \infty $.
In terms of $p$ the field equations (\ref{first}--\ref{last})
are, where $\xi = \frac {1}{\Lambda }$,
\bea
p' & = &  \left( \xi -\frac {2p}{r} -\frac {r^{2}}{3}\right) 
\omega '^{2} +\frac {\xi \left( \omega ^{2}-1 \right) ^{2}}{2r^{2}}
\nn
\\
\frac {S'}{S} & = & \frac {2\omega '^{2}}{r}
\nn
\\ 
0 & = & r^{2} \left( \xi -\frac {2p}{r} -\frac {r^{2}}{3}
\right) \omega '' +\left( 2p-\frac {2r^{3}}{3}-
\frac {\xi \left( \omega ^{2}-1 \right) ^{2}}{r} \right)
\omega ' 
\nn
\\
& & 
+ \xi \left( 1-\omega ^{2}\right) \omega .  
\label{xieqns}
\eea 
When $\xi =0$, we have the solution 
$p(r)=-\frac {\rh ^{3}}{6}$, 
$\omega (r)=\omega _{h}$, $S(r)=1$ for any value of 
$\omega _{h}$.
By a local existence theorem (similar to  Lemma \ref{horexist}), 
the solutions are continuous in the initial values and $\xi $.
Therefore, given a value of $\omega _{h}$, for $\xi $ sufficiently
small, the solution will remain close to its values for  $\xi =0$,
until $r\gg 1$ and $p(r)/r\ll 1/|\Lambda |$.
In particular, $\omega $ will have no nodes in this regime. 
Since $m(r)/r\ll 1$, we may consider the asymptotic equation 
(\ref{asympt}). 
The argument used at the end of the proof of Proposition
\ref{contprop} then shows that we have a solution for all $r$ and
$\omega $ will have no zeros.
\hfill
$\square $

\bigskip
Once again, this proposition does not carry over for $\Lambda >0$.
Quite the opposite may be inferred from 
\cite{review,volkov,coscol}, namely that
non-trivial solutions exist only for sufficiently small values
of $|\Lambda |$.
This result is the crux of our proof that there exist stable
black holes in EYM theory with a negative cosmological constant.
Black holes for which $|\Lambda |\gg 1$ are such that the 
length scale set by the cosmological constant $l$ (given
by $l^{-2}=-\Lambda /3$) is much smaller than the radius of
the event horizon.  This means that we have in effect a 
gauge field perturbation of anti-de Sitter space.
For illustrative purposes, we have used the value 
$\Lambda =-100$ in figures \ref{l100w08}-\ref{l100d},
which gives $l<1=r_{h}$.
However, it should be noted that the same effects occur
for comparatively small values of $|\Lambda |$, for example,
there are black holes for which $\omega $ has no zeros for
$\Lambda =-0.1$ which satisfy the criteria for stability.
Therefore, the analytic results suggest that the stable
configurations are merely perturbations of anti-de Sitter space,
in fact the results hold for black holes in which $l>r_{h}$,
so that the geometry is quite different from AdS.
The thermodynamics of Schwarzschild-AdS \cite{hawking} suggests
that relatively large black holes (having $r_{h}>l$) may in 
fact be the most interesting since they (in the non-hairy case) are
thermodynamically stable.

\section{The existence of stable black holes}
\label{stable}

\subsection{Linearized perturbation equations}

Consider spherically symmetric perturbations of the black holes
described in the previous section.
The metric is now time-dependent but remains spherically symmetric:
\be
ds^{2}=-N(r,t)S^{2}(r,t) \, dt^{2}+N^{-1}(r,t) \, dr^{2}
+r^{2} \, d\theta ^{2} +r^{2} \sin ^{2}\theta \, d\phi ^{2},
\ee
and we consider the following, more general,
ansatz for the gauge field:
\be
A=a_{0}\tau _{r} \, dt + b\tau _{r} \, dr + \left[
\nu \tau _{\theta }-(1+\omega )\tau _{\phi } \right] d\theta
+\left[ (1+\omega )\tau _{\theta } +\nu \tau _{\phi } \right]
\sin \theta \, d \phi ,
\ee
where $a_{0}$, $b$, $\omega $ and $\nu $ depend on $t$ and $r$.
All the field variables will be written  as follows, for example,
\be
\omega (r,t)=\omega (r)+\delta \omega (r,t)
\ee
where $\omega (r)$ is the static equilibrium solution whose stability
we are investigating, and $\delta \omega (r,t)$ is the perturbation.
We shall work to first order only in small perturbations.
In the case where $a_{0}\equiv 0$, the linearized perturbation equations
decouple into two sectors, as in the
$\Lambda =0$ case \cite{arbstab}.
Firstly, the equations for the perturbations $\delta b$ and 
$\delta \nu $ decouple
from the remaining equations to form the {\em {sphaleronic sector}}
(with ${\dot {}}$ denoting $d/dt$):
\bea
0 & = & {\ddot {\delta b}} +\frac {2NS^{2}}{r^{2}} \left[
\omega \left( \delta \nu '+\omega \delta b \right) 
-\omega ' \delta \nu \right]
\label{ympert}
\\
0 & = & {\ddot {\delta \nu }} -NS\left( NS \right) ' \left[
\delta \nu '+\omega \delta b \right] - N^{2} S^{2} \left[
\delta \nu '' +\omega \delta b' +2\omega '\delta b \right]
\nn
\\
& &  
+\frac {NS^{2}}{r^{2}}\delta \nu \left( \omega ^{2}-1 \right) .
\eea
The variables $\delta b$ and $\delta \nu $ must also satisfy the 
{\em {Gauss constraint}} \cite{arbstab}:
\be
\left( \frac {2}{rS^{2}} -\frac {S'}{S^{3}} \right)
{\dot {\delta b}} +\frac {1}{S^{2}} {\dot {\delta b}}' 
-\frac {2}{r^{2}NS^{2}} \omega {\dot {\delta \nu }} =0.
\label{gauss}
\ee
The remaining perturbations, $\delta \omega $, 
$\delta m$ and $\delta S$ form the {\em {gravitational sector}}.

\subsection{Sphaleronic sector}
Volkov et al \cite{volk} used a powerful method to show that the
$n$-th asymptotically flat 
coloured black hole has exactly $n$ unstable modes in this
sector.
The same method will now be used to show that the solutions 
described in section 2 in which the gauge field $\omega $ has
no nodes have no instabilities in this sector.

\bigskip
Firstly define a new variable $\alpha $ by
\be
\alpha = \frac {r^{2}\delta b}{2S},
\ee
in terms of which the Gauss constraint (\ref{gauss}) reads
\be
{\dot {\delta \nu }}=-\frac {NS}{\omega }{\dot {\alpha }}' ,
\ee
so that $\delta \nu =NS\alpha '/\omega $.
As usual, next define a ``tortoise'' co-ordinate $r^{*}$ by
\be
\frac {dr^{*}}{dr}=\frac {1}{NS}.
\ee
The perturbation equation (\ref{ympert}) takes a simple form when we
introduce the quantity $\beta = \alpha /\omega $ and consider 
periodic perturbations 
$\beta (r^{*},t)=e^{i\sigma t}\beta (r^{*})$
\be
\sigma ^{2}\beta =-\frac {d^{2}\beta }{dr^{*2}} +\beta 
\left\{ \frac {NS^{2}}{r^{2}} \left[ 1+\omega ^{2} \right] 
+\frac {2}{\omega ^{2}} \left( \frac {d\omega }{dr^{*}} \right) ^{2}
\right\} ,
\label{sphaleqn}
\ee
where we have used the static field equation (\ref{last}) for 
$\omega $.
This has the form of a standard Schr\"odinger equation.
The potential is positive and regular everywhere in the case where
$\omega $ has no nodes.
As $r\rightarrow r_{h}$, $r^{*}\rightarrow -\infty $,
the potential tends to zero, and in the other asymptotic
regime, $r$, $r^{*}\rightarrow \infty $, the potential
approaches the constant value
\be
-\frac {\Lambda }{3} \left( 1+\omega _{\infty }^{2} \right) 
>0.
\ee
Then a standard theorem of quantum mechanics \cite{messiah}
tells us that there are no bound states for this system, since
the potential is everywhere greater than the lower of its
two asymptotic values. 
In other
words, there are no negative eigenvalues for $\sigma ^{2}$ and
no unstable modes.

\bigskip
If the gauge function $\omega $ has $n$ nodes, then 
we would expect that the method of
Volkov et al \cite{volk} 
could be extended to this case to show that there
are exactly $n$ unstable sphaleronic modes.
This is the situation for asymptotically flat and asymptotically
de Sitter black holes.

\subsection{Gravitational sector}

The perturbations $\delta m$ and $\delta S$ can be eliminated from 
the equations for this sector to leave a single equation for
$\delta \omega (r,t)=e^{i\sigma t}\delta \omega (r^{*})$:
\be
\sigma ^{2}\delta \omega = 
-\frac {d^{2}\delta \omega }{dr^{*2}} +U(r^{*}) \delta \omega
\label{graveqn}
\ee
where the potential $U$ is given by
\be
U=\frac {NS^{2}}{r^{2}}\left\{
3\omega ^{2}-1-4r\omega '^{2} \left[ \frac {1}{r}-\Lambda r -
\frac {\left( 1-\omega ^{2} \right) ^{2}}{r^{3}} \right]
+\frac {8}{r} \omega \omega ' \left( \omega ^{2}-1 \right) \right\} ,
\label{gravpot}
\ee
with $'$ denoting $d/dr$, as previously.
Near the event horizon, as $r\rightarrow \rh $ and $r^{*}\rightarrow
-\infty $, the potential $U\rightarrow 0$.
At infinity, $r \rightarrow \infty $, and $U$ approaches its
asymptotic value:
\be
U\rightarrow -\frac {\Lambda }{3} \left( 3\omega _{\infty }^{2}-1
\right) .
\ee
As for the sphaleronic sector,
a standard theorem \cite{messiah} tells us that the equation
(\ref{graveqn}) will have no bound states if the potential $U$ is
everywhere greater than the lower of its two asymptotic values.
These conditions will be satisfied if
\be 
\omega _{\infty }\ge 1/{\sqrt {3}} 
\qquad  {\mbox {and}}
\qquad 
U\ge 0 
{\mbox { everywhere.}}
\label{uconds}
\ee

\begin{figure}
% GNUPLOT: LaTeX picture
\setlength{\unitlength}{0.240900pt}
\ifx\plotpoint\undefined\newsavebox{\plotpoint}\fi
\sbox{\plotpoint}{\rule[-0.200pt]{0.400pt}{0.400pt}}%
\begin{picture}(1500,900)(0,0)
\font\gnuplot=cmr10 at 10pt
\gnuplot
\sbox{\plotpoint}{\rule[-0.200pt]{0.400pt}{0.400pt}}%
\put(220.0,113.0){\rule[-0.200pt]{292.934pt}{0.400pt}}
\put(220.0,113.0){\rule[-0.200pt]{4.818pt}{0.400pt}}
\put(198,113){\makebox(0,0)[r]{0}}
\put(1416.0,113.0){\rule[-0.200pt]{4.818pt}{0.400pt}}
\put(220.0,198.0){\rule[-0.200pt]{4.818pt}{0.400pt}}
\put(198,198){\makebox(0,0)[r]{10}}
\put(1416.0,198.0){\rule[-0.200pt]{4.818pt}{0.400pt}}
\put(220.0,283.0){\rule[-0.200pt]{4.818pt}{0.400pt}}
\put(198,283){\makebox(0,0)[r]{20}}
\put(1416.0,283.0){\rule[-0.200pt]{4.818pt}{0.400pt}}
\put(220.0,368.0){\rule[-0.200pt]{4.818pt}{0.400pt}}
\put(198,368){\makebox(0,0)[r]{30}}
\put(1416.0,368.0){\rule[-0.200pt]{4.818pt}{0.400pt}}
\put(220.0,453.0){\rule[-0.200pt]{4.818pt}{0.400pt}}
\put(198,453){\makebox(0,0)[r]{40}}
\put(1416.0,453.0){\rule[-0.200pt]{4.818pt}{0.400pt}}
\put(220.0,537.0){\rule[-0.200pt]{4.818pt}{0.400pt}}
\put(198,537){\makebox(0,0)[r]{50}}
\put(1416.0,537.0){\rule[-0.200pt]{4.818pt}{0.400pt}}
\put(220.0,622.0){\rule[-0.200pt]{4.818pt}{0.400pt}}
\put(198,622){\makebox(0,0)[r]{60}}
\put(1416.0,622.0){\rule[-0.200pt]{4.818pt}{0.400pt}}
\put(220.0,707.0){\rule[-0.200pt]{4.818pt}{0.400pt}}
\put(198,707){\makebox(0,0)[r]{70}}
\put(1416.0,707.0){\rule[-0.200pt]{4.818pt}{0.400pt}}
\put(220.0,792.0){\rule[-0.200pt]{4.818pt}{0.400pt}}
\put(198,792){\makebox(0,0)[r]{80}}
\put(1416.0,792.0){\rule[-0.200pt]{4.818pt}{0.400pt}}
\put(220.0,877.0){\rule[-0.200pt]{4.818pt}{0.400pt}}
\put(198,877){\makebox(0,0)[r]{90}}
\put(1416.0,877.0){\rule[-0.200pt]{4.818pt}{0.400pt}}
\put(220.0,113.0){\rule[-0.200pt]{0.400pt}{4.818pt}}
\put(220,68){\makebox(0,0){1}}
\put(220.0,857.0){\rule[-0.200pt]{0.400pt}{4.818pt}}
\put(281.0,113.0){\rule[-0.200pt]{0.400pt}{2.409pt}}
\put(281.0,867.0){\rule[-0.200pt]{0.400pt}{2.409pt}}
\put(362.0,113.0){\rule[-0.200pt]{0.400pt}{2.409pt}}
\put(362.0,867.0){\rule[-0.200pt]{0.400pt}{2.409pt}}
\put(403.0,113.0){\rule[-0.200pt]{0.400pt}{2.409pt}}
\put(403.0,867.0){\rule[-0.200pt]{0.400pt}{2.409pt}}
\put(423.0,113.0){\rule[-0.200pt]{0.400pt}{4.818pt}}
\put(423,68){\makebox(0,0){10}}
\put(423.0,857.0){\rule[-0.200pt]{0.400pt}{4.818pt}}
\put(484.0,113.0){\rule[-0.200pt]{0.400pt}{2.409pt}}
\put(484.0,867.0){\rule[-0.200pt]{0.400pt}{2.409pt}}
\put(564.0,113.0){\rule[-0.200pt]{0.400pt}{2.409pt}}
\put(564.0,867.0){\rule[-0.200pt]{0.400pt}{2.409pt}}
\put(606.0,113.0){\rule[-0.200pt]{0.400pt}{2.409pt}}
\put(606.0,867.0){\rule[-0.200pt]{0.400pt}{2.409pt}}
\put(625.0,113.0){\rule[-0.200pt]{0.400pt}{4.818pt}}
\put(625,68){\makebox(0,0){100}}
\put(625.0,857.0){\rule[-0.200pt]{0.400pt}{4.818pt}}
\put(686.0,113.0){\rule[-0.200pt]{0.400pt}{2.409pt}}
\put(686.0,867.0){\rule[-0.200pt]{0.400pt}{2.409pt}}
\put(767.0,113.0){\rule[-0.200pt]{0.400pt}{2.409pt}}
\put(767.0,867.0){\rule[-0.200pt]{0.400pt}{2.409pt}}
\put(808.0,113.0){\rule[-0.200pt]{0.400pt}{2.409pt}}
\put(808.0,867.0){\rule[-0.200pt]{0.400pt}{2.409pt}}
\put(828.0,113.0){\rule[-0.200pt]{0.400pt}{4.818pt}}
\put(828,68){\makebox(0,0){1000}}
\put(828.0,857.0){\rule[-0.200pt]{0.400pt}{4.818pt}}
\put(889.0,113.0){\rule[-0.200pt]{0.400pt}{2.409pt}}
\put(889.0,867.0){\rule[-0.200pt]{0.400pt}{2.409pt}}
\put(970.0,113.0){\rule[-0.200pt]{0.400pt}{2.409pt}}
\put(970.0,867.0){\rule[-0.200pt]{0.400pt}{2.409pt}}
\put(1011.0,113.0){\rule[-0.200pt]{0.400pt}{2.409pt}}
\put(1011.0,867.0){\rule[-0.200pt]{0.400pt}{2.409pt}}
\put(1031.0,113.0){\rule[-0.200pt]{0.400pt}{4.818pt}}
\put(1031,68){\makebox(0,0){10000}}
\put(1031.0,857.0){\rule[-0.200pt]{0.400pt}{4.818pt}}
\put(1092.0,113.0){\rule[-0.200pt]{0.400pt}{2.409pt}}
\put(1092.0,867.0){\rule[-0.200pt]{0.400pt}{2.409pt}}
\put(1172.0,113.0){\rule[-0.200pt]{0.400pt}{2.409pt}}
\put(1172.0,867.0){\rule[-0.200pt]{0.400pt}{2.409pt}}
\put(1214.0,113.0){\rule[-0.200pt]{0.400pt}{2.409pt}}
\put(1214.0,867.0){\rule[-0.200pt]{0.400pt}{2.409pt}}
\put(1233.0,113.0){\rule[-0.200pt]{0.400pt}{4.818pt}}
\put(1233,68){\makebox(0,0){100000}}
\put(1233.0,857.0){\rule[-0.200pt]{0.400pt}{4.818pt}}
\put(1294.0,113.0){\rule[-0.200pt]{0.400pt}{2.409pt}}
\put(1294.0,867.0){\rule[-0.200pt]{0.400pt}{2.409pt}}
\put(1375.0,113.0){\rule[-0.200pt]{0.400pt}{2.409pt}}
\put(1375.0,867.0){\rule[-0.200pt]{0.400pt}{2.409pt}}
\put(1416.0,113.0){\rule[-0.200pt]{0.400pt}{2.409pt}}
\put(1416.0,867.0){\rule[-0.200pt]{0.400pt}{2.409pt}}
\put(1436.0,113.0){\rule[-0.200pt]{0.400pt}{4.818pt}}
\put(1436,68){\makebox(0,0){1e+06}}
\put(1436.0,857.0){\rule[-0.200pt]{0.400pt}{4.818pt}}
\put(220.0,113.0){\rule[-0.200pt]{292.934pt}{0.400pt}}
\put(1436.0,113.0){\rule[-0.200pt]{0.400pt}{184.048pt}}
\put(220.0,877.0){\rule[-0.200pt]{292.934pt}{0.400pt}}
\put(45,495){\makebox(0,0){$U(r)$}}
\put(828,23){\makebox(0,0){$r$}}
\put(220.0,113.0){\rule[-0.200pt]{0.400pt}{184.048pt}}
\put(220,113){\usebox{\plotpoint}}
\put(220,113){\usebox{\plotpoint}}
\put(220,113){\usebox{\plotpoint}}
\put(220,113){\usebox{\plotpoint}}
\put(220,113){\usebox{\plotpoint}}
\put(220,113){\usebox{\plotpoint}}
\put(220,113){\usebox{\plotpoint}}
\put(220,113){\usebox{\plotpoint}}
\put(220,113){\usebox{\plotpoint}}
\put(219.67,114){\rule{0.400pt}{1.927pt}}
\multiput(219.17,114.00)(1.000,4.000){2}{\rule{0.400pt}{0.964pt}}
\multiput(221.61,122.00)(0.447,5.597){3}{\rule{0.108pt}{3.567pt}}
\multiput(220.17,122.00)(3.000,18.597){2}{\rule{0.400pt}{1.783pt}}
\put(224.17,148){\rule{0.400pt}{3.300pt}}
\multiput(223.17,148.00)(2.000,9.151){2}{\rule{0.400pt}{1.650pt}}
\multiput(226.59,164.00)(0.489,3.116){15}{\rule{0.118pt}{2.500pt}}
\multiput(225.17,164.00)(9.000,48.811){2}{\rule{0.400pt}{1.250pt}}
\multiput(235.58,218.00)(0.499,1.122){115}{\rule{0.120pt}{0.995pt}}
\multiput(234.17,218.00)(59.000,129.935){2}{\rule{0.400pt}{0.497pt}}
\multiput(294.00,350.59)(1.154,0.488){13}{\rule{1.000pt}{0.117pt}}
\multiput(294.00,349.17)(15.924,8.000){2}{\rule{0.500pt}{0.400pt}}
\multiput(312.00,358.59)(13.066,0.477){7}{\rule{9.540pt}{0.115pt}}
\multiput(312.00,357.17)(98.199,5.000){2}{\rule{4.770pt}{0.400pt}}
\put(220.0,113.0){\usebox{\plotpoint}}
\put(476,363.17){\rule{9.100pt}{0.400pt}}
\multiput(476.00,362.17)(26.112,2.000){2}{\rule{4.550pt}{0.400pt}}
\put(521,364.67){\rule{11.081pt}{0.400pt}}
\multiput(521.00,364.17)(23.000,1.000){2}{\rule{5.541pt}{0.400pt}}
\put(430.0,363.0){\rule[-0.200pt]{11.081pt}{0.400pt}}
\put(567.0,366.0){\rule[-0.200pt]{209.342pt}{0.400pt}}
\put(220,113){\usebox{\plotpoint}}
\put(220.00,113.00){\usebox{\plotpoint}}
\multiput(220,114)(2.935,20.547){0}{\usebox{\plotpoint}}
\multiput(221,121)(1.295,20.715){2}{\usebox{\plotpoint}}
\multiput(224,169)(2.032,20.656){3}{\usebox{\plotpoint}}
\multiput(230,230)(6.052,19.854){14}{\usebox{\plotpoint}}
\multiput(312,499)(20.108,5.144){2}{\usebox{\plotpoint}}
\multiput(355,510)(20.750,0.483){2}{\usebox{\plotpoint}}
\multiput(398,511)(20.750,0.483){2}{\usebox{\plotpoint}}
\multiput(441,512)(20.751,0.451){2}{\usebox{\plotpoint}}
\multiput(487,513)(20.756,0.000){2}{\usebox{\plotpoint}}
\multiput(532,513)(20.756,0.000){2}{\usebox{\plotpoint}}
\multiput(576,513)(20.756,0.000){3}{\usebox{\plotpoint}}
\multiput(621,513)(20.756,0.000){2}{\usebox{\plotpoint}}
\multiput(667,513)(20.756,0.000){2}{\usebox{\plotpoint}}
\multiput(712,513)(20.756,0.000){2}{\usebox{\plotpoint}}
\multiput(757,513)(20.756,0.000){2}{\usebox{\plotpoint}}
\multiput(801,513)(20.756,0.000){3}{\usebox{\plotpoint}}
\multiput(846,513)(20.756,0.000){2}{\usebox{\plotpoint}}
\multiput(891,513)(20.756,0.000){2}{\usebox{\plotpoint}}
\multiput(937,513)(20.756,0.000){2}{\usebox{\plotpoint}}
\multiput(981,513)(20.756,0.000){2}{\usebox{\plotpoint}}
\multiput(1025,513)(20.756,0.000){2}{\usebox{\plotpoint}}
\multiput(1072,513)(20.756,0.000){3}{\usebox{\plotpoint}}
\multiput(1117,513)(20.756,0.000){2}{\usebox{\plotpoint}}
\multiput(1162,513)(20.756,0.000){2}{\usebox{\plotpoint}}
\multiput(1208,513)(20.756,0.000){2}{\usebox{\plotpoint}}
\multiput(1254,513)(20.756,0.000){2}{\usebox{\plotpoint}}
\multiput(1298,513)(20.756,0.000){2}{\usebox{\plotpoint}}
\multiput(1343,513)(20.756,0.000){3}{\usebox{\plotpoint}}
\multiput(1389,513)(20.756,0.000){2}{\usebox{\plotpoint}}
\multiput(1434,513)(20.756,0.000){0}{\usebox{\plotpoint}}
\put(1436,513){\usebox{\plotpoint}}
\sbox{\plotpoint}{\rule[-0.400pt]{0.800pt}{0.800pt}}%
\put(220,113){\usebox{\plotpoint}}
\put(220,113){\usebox{\plotpoint}}
\put(220,113){\usebox{\plotpoint}}
\put(220,113){\usebox{\plotpoint}}
\put(220,113){\usebox{\plotpoint}}
\put(220,113){\usebox{\plotpoint}}
\put(220,113){\usebox{\plotpoint}}
\put(220,113){\usebox{\plotpoint}}
\put(220,113){\usebox{\plotpoint}}
\put(220,113){\usebox{\plotpoint}}
\put(220,113){\usebox{\plotpoint}}
\put(218.84,116){\rule{0.800pt}{5.300pt}}
\multiput(218.34,116.00)(1.000,11.000){2}{\rule{0.800pt}{2.650pt}}
\put(220.34,138){\rule{0.800pt}{9.636pt}}
\multiput(219.34,138.00)(2.000,20.000){2}{\rule{0.800pt}{4.818pt}}
\put(221.84,178){\rule{0.800pt}{9.395pt}}
\multiput(221.34,178.00)(1.000,19.500){2}{\rule{0.800pt}{4.698pt}}
\multiput(225.41,217.00)(0.501,2.323){267}{\rule{0.121pt}{3.902pt}}
\multiput(222.34,217.00)(137.000,625.901){2}{\rule{0.800pt}{1.951pt}}
\multiput(361.00,852.40)(3.278,0.526){7}{\rule{4.657pt}{0.127pt}}
\multiput(361.00,849.34)(29.334,7.000){2}{\rule{2.329pt}{0.800pt}}
\put(400,857.34){\rule{4.336pt}{0.800pt}}
\multiput(400.00,856.34)(9.000,2.000){2}{\rule{2.168pt}{0.800pt}}
\put(418,858.84){\rule{7.709pt}{0.800pt}}
\multiput(418.00,858.34)(16.000,1.000){2}{\rule{3.854pt}{0.800pt}}
\put(450,860.34){\rule{12.045pt}{0.800pt}}
\multiput(450.00,859.34)(25.000,2.000){2}{\rule{6.022pt}{0.800pt}}
\put(500,861.84){\rule{12.045pt}{0.800pt}}
\multiput(500.00,861.34)(25.000,1.000){2}{\rule{6.022pt}{0.800pt}}
\put(550,862.84){\rule{4.336pt}{0.800pt}}
\multiput(550.00,862.34)(9.000,1.000){2}{\rule{2.168pt}{0.800pt}}
\put(220.0,113.0){\usebox{\plotpoint}}
\put(568.0,865.0){\rule[-0.400pt]{209.101pt}{0.800pt}}
\end{picture}
\caption{The potential $U(r)$ arising in the gravitational
perturbation equations (\ref{graveqn}) and (\ref{gravpot}), 
plotted for the black holes of figures \ref{l100w08},
\ref{l100w09} and \ref{l100w11}.
The key fact here is that $U$ is everywhere positive, so that
these solutions are stable in the gravitational sector.
}
\label{l100u}
\end{figure}
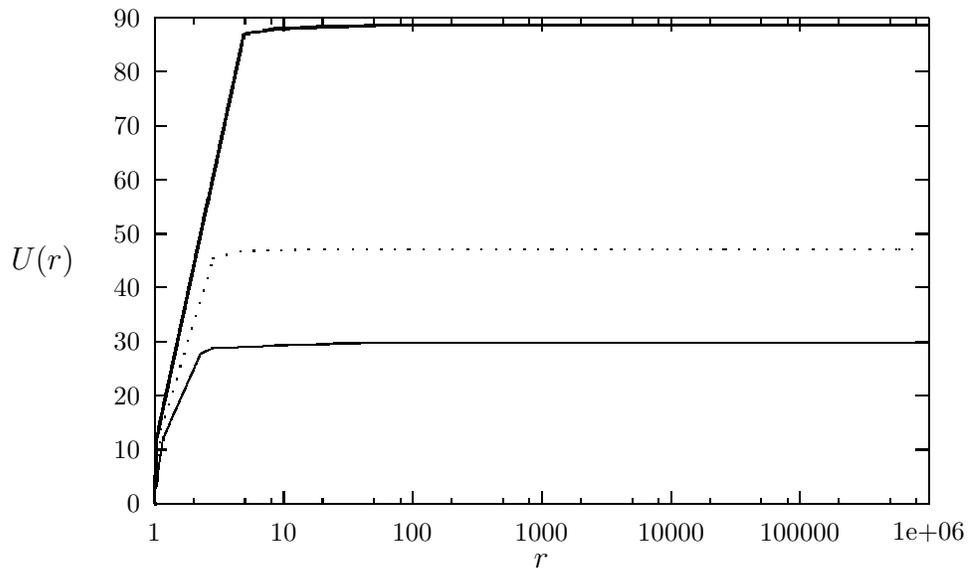

\bigskip
We are particularly interested in those solutions in which the 
gauge field has no zeros, since these are stable in the sphaleronic 
sector.
Figure \ref{l100u} illustrates numerically that there are
such black holes which also satisfy our two requirements for
stability in the gravitational sector.
These solutions are therefore linearly stable in both sectors 
of the theory. 

\bigskip
We shall now proceed to show analytically that there exist
black holes in which the gauge field has no zeros and the
gravitational potential satisfies the conditions
(\ref{uconds}).
Firstly, we restrict attention to those solutions
for which $\omega _{h}>1/{\sqrt {3}}$.
Then, from proposition \ref{largeprop}, for
$|\Lambda |$ sufficiently large, $\omega $ will remain
sufficiently close to its initial value,
so that $\omega (r)>1/{\sqrt {3}}$ for all $r$,
 and in particular,
$\omega _{\infty }>1/{\sqrt {3}}$.
In this case, $\omega '$ will be very small for all $r$,
and the dominant behaviour of the potential $U$ will be:
\be
U=\frac {NS^{2}}{r^{2}} \left[
3\omega ^{2}-1+4\Lambda r^{2}\omega '^{2} \right] .
\ee
The third term in the brackets is negative, whereas the 
other two terms together are positive.
If we can show that this third term can be made as small as
we like by taking $|\Lambda |$ sufficiently large, then
$U>0$ for all $r^{*}$ and the black hole has no unstable
modes in the gravitational sector.
Therefore, we need to show that 
${\sqrt {|\Lambda |}} \omega '\rightarrow 0$ as
$|\Lambda |\rightarrow \infty $.

\bigskip
Firstly, consider the expression (\ref{omegah}) for
$\omega '(r_{h})$:
\be
\omega _{h}'=\frac {(\omega _{h}^{2}-1)\omega _{h}}{r_{h}
-\Lambda r_{h}^{3}-
\frac {(\omega _{h}^{2}-1)^{2}}{r_{h} } }
\sim
-\frac {1}{\Lambda r_{h}^{3}}\left( \omega _{h}^{2}-1 \right)
\omega _{h} 
\label{omegaph}
\ee
as $|\Lambda |\rightarrow \infty $.
From (\ref{omegaph}), it is clear that 
${\sqrt {|\Lambda |}}\omega _{h}'\rightarrow 0$ as
$|\Lambda |\rightarrow \infty $.
In order to show that this is true for other values of
$r$ also, we turn to the equations (\ref{xieqns})
for the field variables, where $\xi =1/\Lambda $
and $p(r)=m(r)/\Lambda $.
Introduce a new variable $\eta (r)$ by
$\eta (r)=\omega '(r)/{\sqrt {|\xi |}}$, in terms of which
the equations (\ref{xieqns}) read:
\bea
p' & = & 
\left( 
\xi -\frac {2p}{r} -\frac {r^{2}}{3} \right) 
|\xi | \eta ^{2}+
\frac {\xi }{2r^{2}} \left( \omega ^{2}-1 \right) ^{2}
\nn
\\
\frac {S'}{S} & = & \frac {2|\xi |\eta ^{2}}{r}
\nn
\\
0 & = &
r^{2} \left( \xi -\frac {2p}{r} -\frac {r^{2}}{3} \right)
\eta '+\left(
2p - \frac {\xi }{r} \left( \omega ^{2}-1 \right) ^{2}
-\frac {2r^{3}}{3} \right) \eta
\nn \\
& & 
+{\sqrt {|\xi |}} \left( 1-\omega ^{2} \right) \omega .
\eea
When $\xi =0$, we have the solution $p\equiv -r_{h}^{3}/6$,
$\omega \equiv \omega _{h}$, $S\equiv 1$ and the 
equation for $\eta $ decouples from the rest to give
\be
r\left( r_{h}^{3}-r^{3} \right) \eta '
-\left( r_{h}^{3}+2r^{3} \right) \eta =0.
\ee
This is a linear first order differential equation with solution
\be
\eta (r)=\frac {{\cal {A}}r}{r^{3}-r_{h}^{3}},
\ee
where ${\cal {A}}$ is a constant of integration.
However, we know that $\eta $ must vanish at the event horizon
from (\ref{omegaph}), and therefore ${\cal {A}}=0$,
which means that $\eta $ vanishes identically.
Therefore ${\sqrt {|\Lambda |}}\omega '\rightarrow 0$
as $|\Lambda |\rightarrow \infty $, and the potential 
$U$ is positive for sufficiently large $|\Lambda |$.
This proves that the black holes for which 
$\omega _{h}>1/{\sqrt {3}}$ have no unstable modes in the
gravitational sector for sufficiently large $|\Lambda |$.
These black holes will have a gauge field which has
no zeros and will therefore be stable in both the
gravitational and sphaleronic sectors.

\section{Conclusions}
\label{conc}

In this paper we have studied the ${\mathfrak {su}}(2)$
Einstein-Yang-Mills equations and looked for spherically symmetric 
black hole solutions which approach anti-de Sitter space at infinity.
The numerical integration of the field equations yielded behaviour
rather different from that observed for black holes in asymptotically
flat or asymptotically de Sitter spacetime.
For example, there are regular black hole solutions for continuous
intervals of the initial parameter space, rather than discrete points;
there are solutions for which the gauge field has no zeros;
solutions exist for all values of $\Lambda <0$; and the
gauge field $\omega $ does not necessarily satisfy $|\omega |<1$.
Analysis of the field equations has confirmed this observed behaviour.
In particular, we showed that the solutions in which the
gauge field has no zeros  are (linearly) {\em {stable}}.

\bigskip
The results of this paper may also have consequences for the 
``no-hair'' conjecture.
It is already suspected \cite{torii} that the asymptotic nature
of the spacetime can affect the existence of hair.
In \cite{torii} it is shown that the ``no-scalar-hair'' theorem
of Bekenstein \cite{bek} is no longer valid for asymptotically
de Sitter space, and specific solutions exhibited.
However, although these solutions evade the ``letter'' of the
no-hair conjecture, they do not alter its ``spirit'' since they
are unstable.
The fact that we have found stable solutions in EYM theory 
when the cosmological constant is negative suggests that
it may be possible to find primary scalar hair in asymptotically
anti-de Sitter geometries.
We hope to return to this question in the near future.

\bigskip
The behaviour of a quantum field on a non-trivial
black hole background geometry 
with a negative cosmological constant remains
an interesting open question, since it incorporates both hair (which
itself has radical effects on quantum field theory \cite{mw,decoh})
and anti-de Sitter space (which renders Schwarzschild black holes 
thermodynamically stable \cite{hawking}).
The stability of the solutions we have found means that a study of
the quantum field theory would have more wide-ranging implications,
particularly since interesting effects have been observed in
theories with a negative cosmological constant 
(see, for example \cite{hawk}).
It may also be relevant to some of the deep issues being tackled
presently in the light of Maldacena's \cite{maldacena} conjecture
that thermodynamics of quantum gravity with a negative cosmological
constant is equivalent to the large $N$ thermodynamics of 
standard quantum field theory. 
We hope to return to this question in a subsequent publication.

\section*{Acknowledgements}
We would like to thank Dr. N. E. Mavromatos for helpful discussions.
This work is supported by a fellowship at Oriel College, Oxford.

\begin{figure}
% GNUPLOT: LaTeX picture
\setlength{\unitlength}{0.240900pt}
\ifx\plotpoint\undefined\newsavebox{\plotpoint}\fi
%
\caption{The same solutions as figures \ref{l001w} and \ref{l001m},
but with the metric function $\log S (r)$ plotted.}
\label{l001d}
\end{figure}

\end{document}